\documentclass{PoS}

\usepackage{amsmath}
\usepackage{tikz}

\title{Phase diagram of the two-dimensional O(3) model
from dual lattice simulations}

\ShortTitle{Phase diagram of the O(3) model
from dual lattice simulations}

\author{\speaker{Falk Bruckmann}
        \thanks{supported by DFG (BR 2872/6-1)}\\
        Universit\"at Regensburg, Institut f\"ur Theoretische Physik, 
        Universit\"atsstr.\ 31, 93053
        Regensburg, Germany\\
        E-mail: \email{falk.bruckmann@ur.de}}
        
\author{Christof Gattringer and Thomas Kloiber
        \thanks{supported by FWF (DK W1203-N16, I 1452-N27) and partly by  DFG (SFB TR55)}\\
        Universit\"at Graz, Institut f\"ur Physik, Universit\"atsplatz 5, 8010 Graz, Austria\\
        E-mail: \email{christof.gattringer@uni-graz.at}, \email{thomas.kloiber@uni-graz.at}} 
        
\author{Tin Sulejmanpasic
        \thanks{supported by DOE (E-SC0013036)}\\
        Department of Physics, North Carolina State University, Raleigh, NC, 27695\\
        E-mail: \email{tin.sulejmanpasic@gmail.com}}

\abstract{We have simulated the asymptotically free two-dimensional O(3) model at nonzero chemical potential using the model's dual representation. We first demonstrate how the latter solves the sign (complex action) problem. The system displays a crossover at nonzero temperature, while at zero temperature it undergoes a quantum phase transition when $\mu$ reaches the particle mass (generated dynamically similar to QCD). The density follows a square root behavior universal for repulsive bosons in one spatial dimension. We have also measured the spin stiffness, known to be sensitive to the spatial correlation length, using different scaling trajectories to zero temperature and infinite size. It points to a dynamical critical exponent $z=2$. Comparisons to thermodynamic Bethe ansaetze are shown as well.}

\FullConference{34th annual International Symposium on Lattice Field Theory\\
		24-30 July 2016\\
		University of Southampton, UK}

\begin{document}

\section{The O(3) model at nonzero chemical potential}

In 1+1 dimensions, the O(3) sigma model with its simple kinetic action
\begin{equation}
 S=\frac{1}{2g_0^2}\int\!d{\,^2}x\,(\partial_\nu n_a)^2\,,\quad 
 n_a^{\,2}=1\,,\quad 
 a=1,2,3 \qquad
 \text{(sum convention)}
\end{equation}
for the constrained vector field $n$
shares many properties with QCD. Most prominently these are asymptotic freedom and dimensional transmutation (the coupling $g_0$ is dimensionless as in 3+1 dimensional Yang-Mills theory) giving a mass $m$ to the low energy particle triplet, which is proportional to the system's strong scale. 
Other si\-mi\-la\-ri\-ties to gauge theories are nontrivial topology, instantons (including those of fractional charge), renormalons and the possibility to introduce a $\Theta$-angle. The lattice discretization of the system is known as the Heisenberg model and will be used below. The existence of a global O(3) symmetry results in conserved Noether currents and the corresponding conserved charges, to which a chemical potential $\mu$ can be coupled. Using the third generator of the symmetry group that rotates $n_{1,2}$ into each other, the continuum action becomes~\cite{Hasenfratz:1990zz}:
\begin{equation}
  S=\frac{1}{2g_0^2}\int\!d^2x\,\big[\,
  (\partial_\nu  n_a)^2+
  2i\mu\,(n_1\partial_0 n_2-n_2\partial_0 n_1)
  \underbrace{-\mu^2 (n_1^2+n_2^2)}_{\textstyle =+\mu^2(n_3^2-1)}
  \,\big]\,.
\end{equation}
The new term linear in $\mu$ is imaginary, such that the path integral weight $\exp(-S)$ is not positive definite and one faces the complex action/sign problem\footnote{The time-reversed $n$-configuration has the complex conjugated action, thus adding both configurations under the path integral leads to a real weight, which, however, is not of fixed sign (as in many other systems at $\mu\neq 0$).} that hampers numerical simulations, again as in QCD with chemical potential. The term quadratic in $\mu$ is typical for bosonics systems\footnote{The chemical potential modifies the time derivative to $\partial_0-i\mu T^3$, which appears quadratically, see, e.g., \cite{Bruckmann:2014sla}.}. It suppresses the perpendicular component $n_3$ such that at large $\mu$ the system becomes planar, i.e., similar to the O(2) model. The latter contains vortices that are tightly connected to a Berezinskii-Kosterlitz-Thouless (BKT) transition \cite{Berezinsky:1970fr,Kosterlitz:1973xp}.

We first demonstrate how the sign problem is solved with the help of dual variables, which also works for the O(N) and CP(N-1) series the O(3) model is the lowest member of \cite{Bruckmann:2015sua}\footnote{For a $\mu=0$ implementation and efficiency test of this algorithm in CP(N-1) see \cite{Rindlisbacher:2016cpj}.}. Secondly, the phase diagram of the O(3) model -- motivated by the search for the BKT transition \cite{Bruckmann:2014sla} -- will be analyzed through numerical simulations and analytical tools related to integrability \cite{Bruckmann:2016txt}.

\section{Dual variables: $\mu$-dependence and solution of the sign problem}

We start by writing down the lattice action with chemical potential in spherical coordinates in $n$-space, i.e., $n_1+in_2=\sin\vartheta\exp(i\phi)$, $ n_3=\cos\vartheta$: 
\begin{equation}
\begin{split}
 S=-J\, \sum_{x,\nu}\big(
 &\cos\vartheta(x)\cos\vartheta(x+\hat{\nu})\\
 &+\tfrac{1}{2}\,\sin\vartheta(x)\sin\vartheta(x+\hat{\nu})
 \{
 e^{\,i(\phi(x+\hat{\nu})-\phi(x))}e^{\,\mu \delta_{\nu,0}}
 +e^{-i(\phi(x+\hat{\nu})-\phi(x))}e^{-\mu \delta_{\nu,0}}\}
 \big)\,.
\end{split}  
\label{eq:three}
\end{equation}
As in the Heisenberg model, the action consists of hopping terms in all $\nu$-directions. We have set the lattice spacing to unity, and the continuum limit is reached by sending the (positive) coupling $J$ to infinity. The chemical potential enters in the exponential form affecting the temporal ($\nu=0$) hopping terms for $\phi$ in such a way that the action becomes complex\footnote{unless $\mu$ vanishes or is purely imaginary}. 

The dual variable method consists of two main steps:
%. Firstly,  
\begin{itemize}
 \item[(i)] for each term $S_A$ in the action, expand the weight $\exp(-S_A)$ in a power series with a dual variable $k_A$ as summation index,
 \item[(ii)] integrate out the original fields,
\end{itemize}
obtaining an \textbf{exact representation} of the (lattice) partition function. It is as a sum over the nonnegative integer-valued dual variables with weights containing (from the expansion of the exponentials) powers of the coupling constant divided by factorials suppressing large values of the dual variables. The remaining weight stems from the integration (ii), it depends on the specific physical system, but also on the way it is decomposed into summands $S_A$ in step (i)\footnote{For example, turning the exponentials in Eq.~\eqref{eq:three} into trigonometric functions and dualizing these, the resulting dual representation is much less intutive than the one used here.}.

In our system the superindex $A$ consist of bond indices $(x,\nu)$ and internal indices for the various terms in Eq.~\eqref{eq:three}. We will use dual variables $k^{(1,2)}_\nu(x)$ for the two summands in the second line and focus on the $\phi$-integrals. This part is effectively a dualization of the O(2) or XY model, which has been discussed in \cite{Chandrasekharan:2008gp} and simulated in \cite{Banerjee:2010kc}. For integrating over $\phi$ at some site $x$, we have to collect all the $e^{-i\phi(x)}$-factors from the action: they occur with powers of $k^{(1)}_\nu(x)$ from the first summand, but also with the shifted $k^{(1)}_\nu(x-\hat{\nu})$ with negative powers, in both cases for all directions $\nu$. The second summand contributes powers of $k^{(2)}$ with just the opposite sign. The resulting integral only contains the difference $k^{(1)}_\nu(x)-k^{(2)}_\nu(x)\equiv m_\nu(x)\in\mathbb{Z}$,
\begin{equation}
 \forall x: \: 
 \int_0^{2\pi}\!\!d\phi(x)\, 
 \exp\big(-i\phi(x)
 \sum_\nu\big\{
 m_\nu(x)
 -
 m_\nu(x-\hat{\nu})
 \big\}
 \big)
 =2\pi \delta_{\text{Kronecker}}
 (\nabla_\nu m_\nu(x),0)~,
 \label{eq:four}
\end{equation}
where $\nabla_\nu$ is the discretized divergence, $\nabla_\nu m_\nu(x)=\sum_\nu\{ m_\nu(x)-m_\nu(x-\hat{\nu})\}$. The conservation of the symmetry current $m_\nu$ is thus manifest in this representation, actually for all admissible configurations in the path integral (in contrast to the Noether current, in the derivation of which the equations of motion are used). Consequently, the field variable $m_\nu(x)$ must form closed loops, either as vacuum bubbles or as loops closing around the boundary. The corresponding charge $Q=\sum_{x_1}m_0(x)$ is conserved\footnote{Discretized derivatives do not obey the Leibniz rule, which, however, does not enter the derivation of the conservation equation $Q|_{x_0}-Q|_{x_0-1}=-\sum_{x_1}\{m_1(x)-m_1(x-\hat{1})\}=0$.} and, therefore, can be read off as a net flux through any time slice of fixed $x_0$. Only $m$-loops closing around the temporal boundary contribute to the charge and have the 
interpretation of particle worldlines. In Ref.~\cite{Bruckmann:2015hua} we made use of this picture to extract two-particle wave functions and corresponding phase shifts. 
  
The chemical potentials enters in a way very similar to $\phi$,
\begin{equation}
 \prod\limits_x\, 
 \exp\big(\mu\, \delta_{\nu,0}
 \sum_\nu
 m_\nu
 (x)\big)
 =\exp\big(\mu\sum_x m_0(x)\big)
 =\exp\big(\mu N_t\,\sum_{x_1} m_0(x)\big)
 =\exp\big(\frac{\mu}{T}\,Q\big)~.
 \label{eq:five}
\end{equation}
The coupling of $\mu$ to the conserved charge in the exponential (including the temperature $T$) is just the same as in the defining energy representation of the grand canonical ensemble, but different from the original $n$-field representation we started with. 

Concerning the sign problem one sees immediately that, if there was no such problem at $\mu=0$ -- and indeed all terms in the weight at $\mu=0$ are positive, see \cite{Bruckmann:2016txt} for the full result including the $\vartheta$-integration -- then $\mu$ does not introduce a sign problem either. Therefore, we were able to perform numerical simulations for the thermodynamics of the O(3) model using a combination of conventional local Metropolis updates and worm algorithms for the unconstrained and constrained dual variables, respectively. As a check we have calculated the expectation values of the energy density and the mass-gap as functions of the coupling $J$ which nicely agree with strong and weak coupling expansions \cite{Bruckmann:2016txt}. Numerical data shown below have been obtained for a fixed coupling $J=1.3$, which we expect to be large enough to reflect continuum physics. Dimensionful quantities are normalized with the mass $m$, e.g., our lattice spacing is $a=0.22/m$. 

\section{Crossover at nonzero $T$ and comparison to Bethe ans\"atze}

\begin{figure}
 \includegraphics[width=0.49\linewidth,type=pdf,ext=.pdf,read=.pdf]{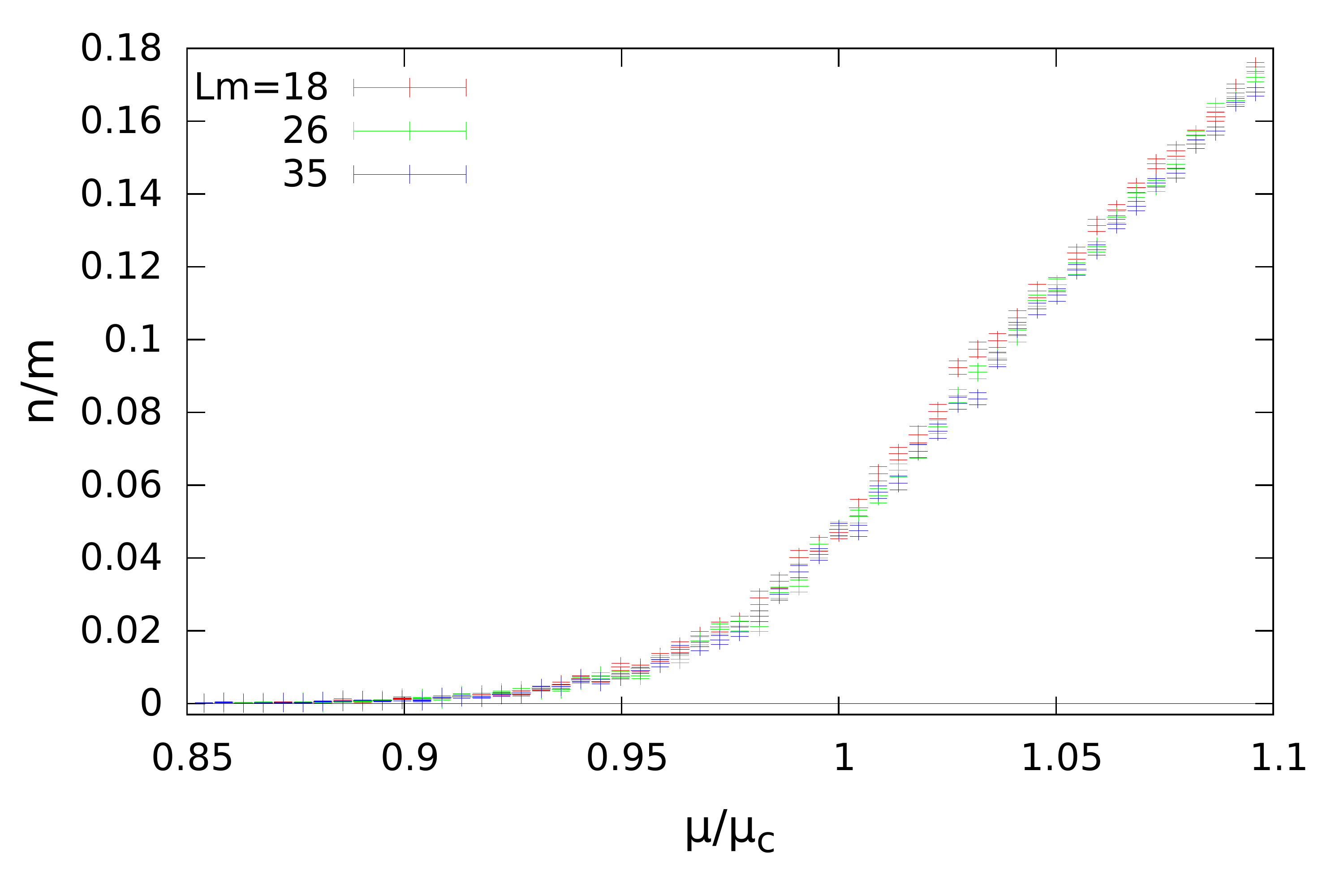}
 \hfill
 \includegraphics[width=0.49\linewidth,type=pdf,ext=.pdf,read=.pdf]{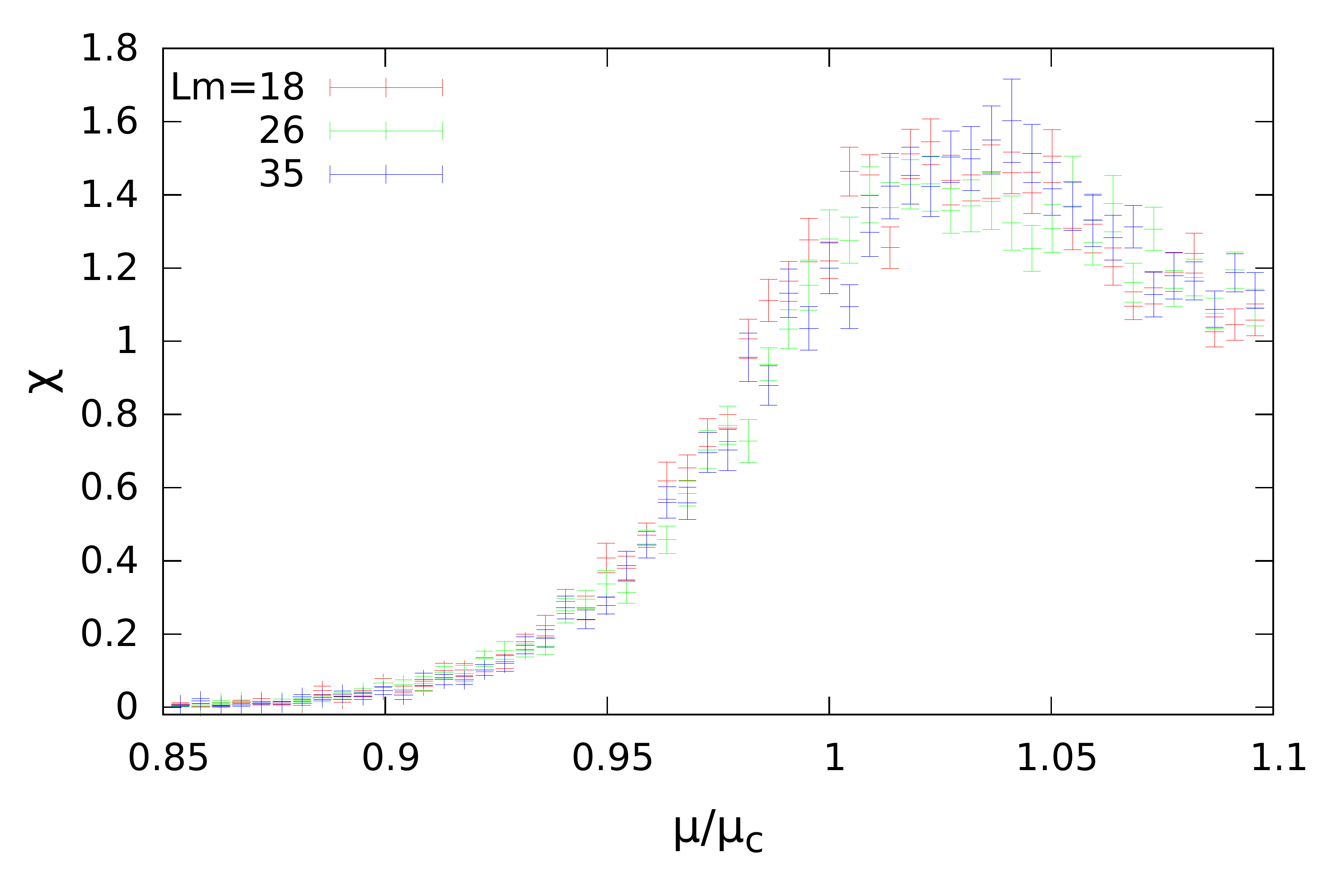}
 \caption{Particle number density $n$ (left) and its susceptibility $\chi_n$ (right) at a low temperature $T = 0.023m$.}
 \label{fig:eins}
\end{figure}

The most prominent thermodynamic quantities are the charge or particle number density,
\begin{equation}
 n=\frac{T}{L}\,\frac{\partial \log Z}{\partial\mu}~,
\end{equation}
and its susceptibility $\chi=\partial n/\partial \mu$. Fig.~\ref{fig:eins} shows our numerical results for them 
at low temperature and three spatial sizes $L$. Both quantities start to increase near a critical value $\mu_c=m$, in line with the interpretation of the chemical potential as the energy to induce particles\footnote{In a similar manner, the second critical $\mu_c$ (in a finite volume) can be used to determine the energy of two particles and thus phase shifts \cite{Bruckmann:2015hua}.} into the system. At nonzero temperatures this increase is smooth and does not scale with the size: the transition is a crossover.

\begin{figure}[t]
 \includegraphics[width=0.49\linewidth,type=pdf,ext=.pdf,read=.pdf]{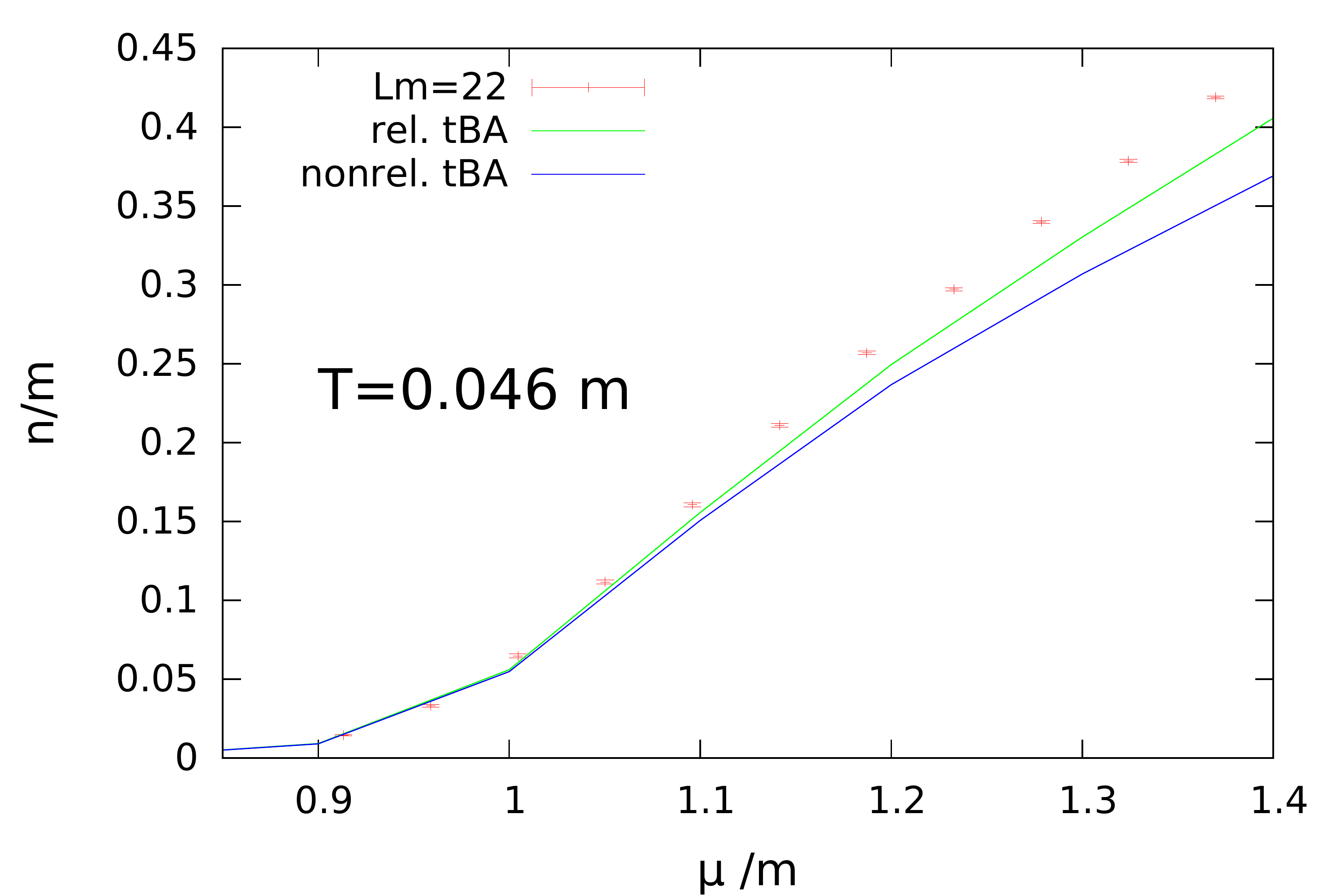}
 \hfill
 \includegraphics[width=0.49\linewidth,type=pdf,ext=.pdf,read=.pdf]{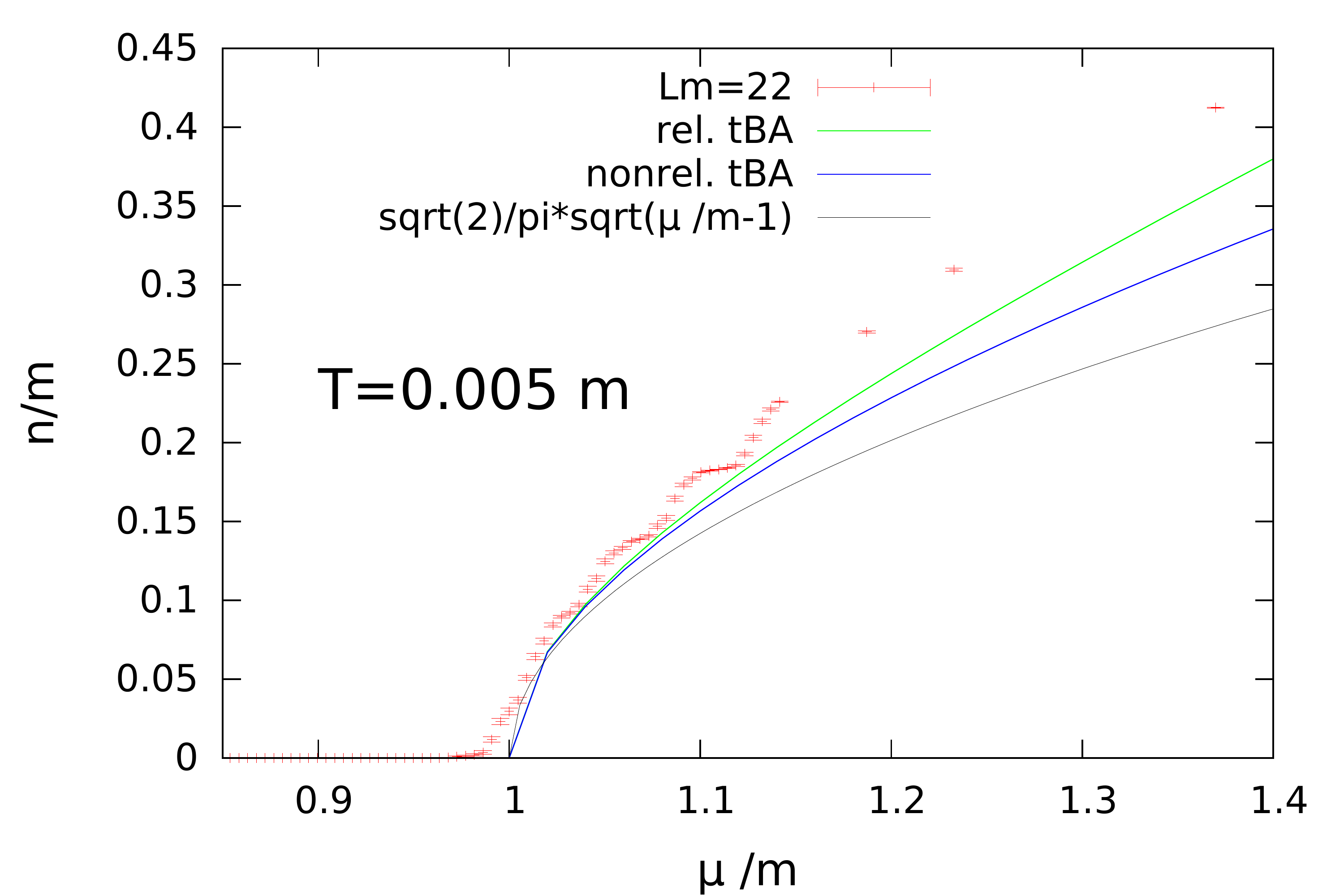}
 \caption{Comparison of low temperature lattice data to relativistic and nonrelativistic thermodynamic Bethe ans\"atze. The black line in the right panel shows the universal square root of Eq.~\protect\eqref{eq:six}.}
 \label{fig:zwei}
\end{figure}

The induced particles are subject to a short range repulsion with known phase shifts \cite{Zamolodchikov:1977nu,Hasenfratz:1990zz}. For low relative momenta, i.e., in the infrared, the system resembles the Lieb-Liniger (LL) model \cite{Lieb:1963rt}, where one-dimensional nonrelativistic bosons repel through a Dirac delta potential and the wave functions consist of plane waves. Such Bethe ans\"atze have been extended to thermodynamics \cite{Yang:1968rm} and  should also describe the O(3) model at low densities. As Fig.~\ref{fig:zwei} shows, these ans\"atze compare well to our numerical data, although they do not contain antibosons.
In particular, for very low momenta the phase shift is universal, and the system's eigenvalues are that of one-dimensional free fermions (eigenfunctions are still symmetric under permutation of the bosons). The density follows immediately from the Fermi-Dirac integral as 
\begin{equation}
 \frac{n}{m}\stackrel{T\to 0}{\longrightarrow}
 \frac{\sqrt{2}}{\pi}\sqrt{\mu/m-1}\cdot\Theta_{\text{Heaviside}}(\mu-m)~.
 \label{eq:six}
\end{equation}
Our numerical data approach this universal curve for very low temperatures, see Fig.~\ref{fig:zwei} right panel, including the vanishing of the density for $\mu$ below the mass threshold for $T=0$. This suggests a second order phase transition at zero temperature, to be discussed in the next section.

\section{Quantum phase transition at $T=0$\,, scaling trajectories and spin stiffness}

In the thermodynamic limit at zero temperature, the extensions of space, $L$, and of Euclidean time, $\beta$, are sent to infinity, and so are the number of lattice sites $N_s$ and $N_t$. One possibility to achieve this is to send $L\to\infty$ at fixed $T$ -- which gives crossovers (see above) -- and to send $T\to 0$ afterwards. For the one-dimensional fermions argued for above the crossover curves become sharper and in this way approach the square root in Eq.~\eqref{eq:six}. From the fermions one can also learn that in the opposite case, first $T\to0$ at fixed $L$, then $L\to\infty$, the square root is approached by an ever finer staircase function \cite{Bruckmann:2016txt}. However, such consecutive limits are computationally expensive. We prefer to send both $T$ and $L^{-1}$ to zero along trajectories characterized by a nonnegative parameter~$\alpha$:
\bigskip

%
%\begin{figure}[!b]\centering
\begin{minipage}{0.35\linewidth}
\begin{equation}
 \begin{split} 
  &T, L^{-1}\to 0\\
  &T L^\alpha \text{ fixed}
 \end{split}
 \label{eq:sixprime}
\end{equation}
\end{minipage}\hfill
\begin{minipage}{0.45\linewidth}
 \begin{tikzpicture}[scale=0.67]
 \fill[black] (0,0) circle (0.15);
 \draw[<-, line width=1pt] (0,0.3) -- (0,4) -- (3.2,4);
 \draw[->, line width=1pt] (0,4.2) -- (0,5);
 \draw[<-, line width=1pt] (0.3,0) -- (4,0) -- (4,3.2);
 \draw[->, line width=1pt] (4.2,0) -- (5,0);
 \draw[<-, line width=1pt] (0.3,0.3) -- (3.4,3.4);
 \draw[<-, line width=1pt] (0.3,0.05) .. controls (2.2,0.25) and (3,0.5) ..  (3.85,3.2);
 \node (T) at (-0.6,4.7) {$T$};
 \node (L) at (5.7,0) {$L^{-1}$};
 \node (a0) at (1.9,4.45) {$\alpha=0$};
 \node (a1) at (1.9,2.45) {$1$};
 \node (a2) at (2.6,1.5) {$2$};
 \node (ainf) at (4.5,1.5) {$\infty$};
 \end{tikzpicture}
\end{minipage}
\bigskip

\noindent For the practical implementation we fix $(T/m)(Lm)^\alpha$ to a dimensionless constant. While the two cases above amount to the extremes $\alpha=0$ and $\alpha=\infty$
(see the sketch of the trajectories in Eq.~\eqref{eq:sixprime}),
in the remainder we will analyse the scaling trajectories $\alpha=1$, i.e., square lattices, and $\alpha=2$, i.e., time elongated lattices.

\begin{figure}[!t]
 \includegraphics[width=0.49\linewidth,type=pdf,ext=.pdf,read=.pdf]{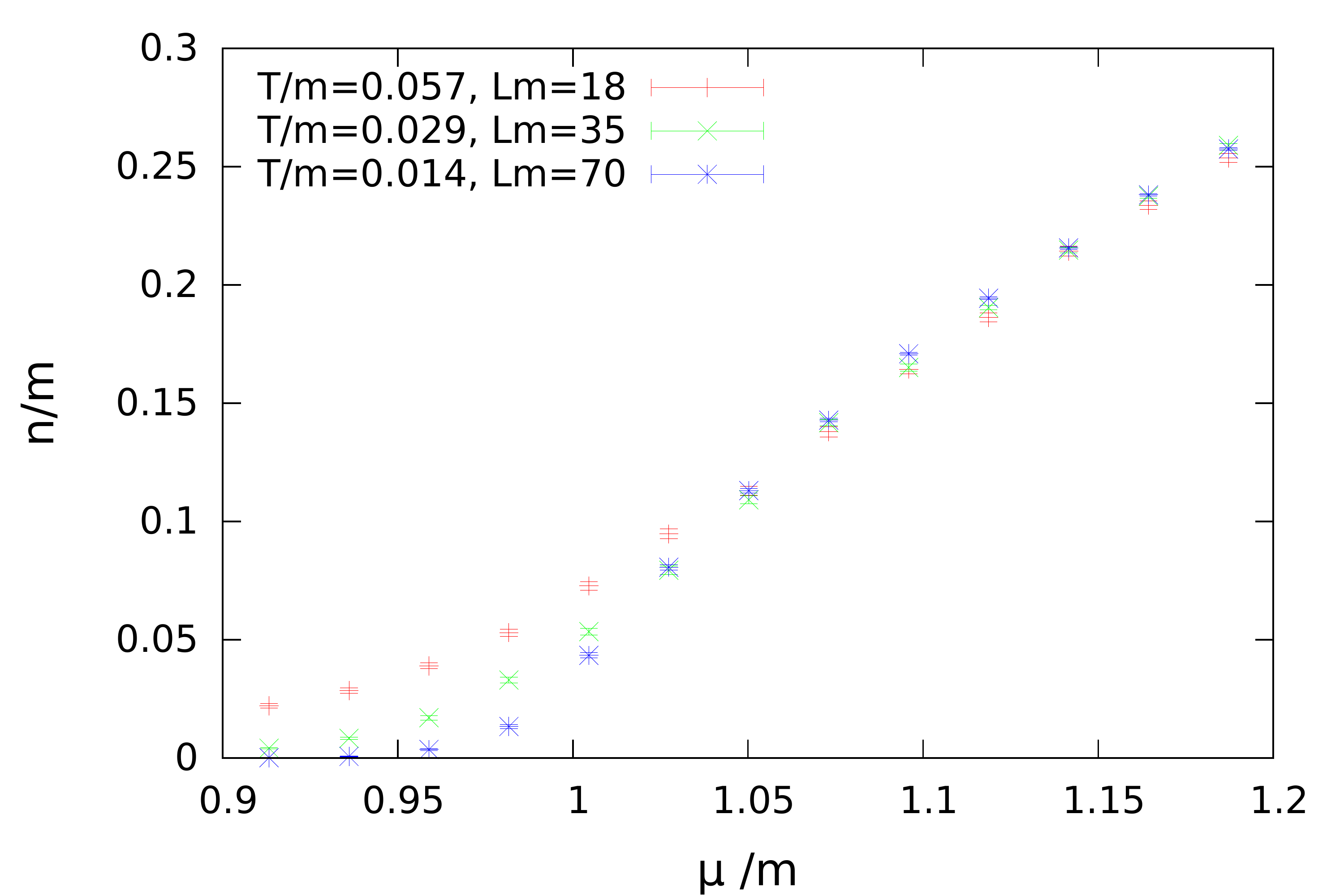}
 \hfill
 \includegraphics[width=0.49\linewidth,type=pdf,ext=.pdf,read=.pdf]{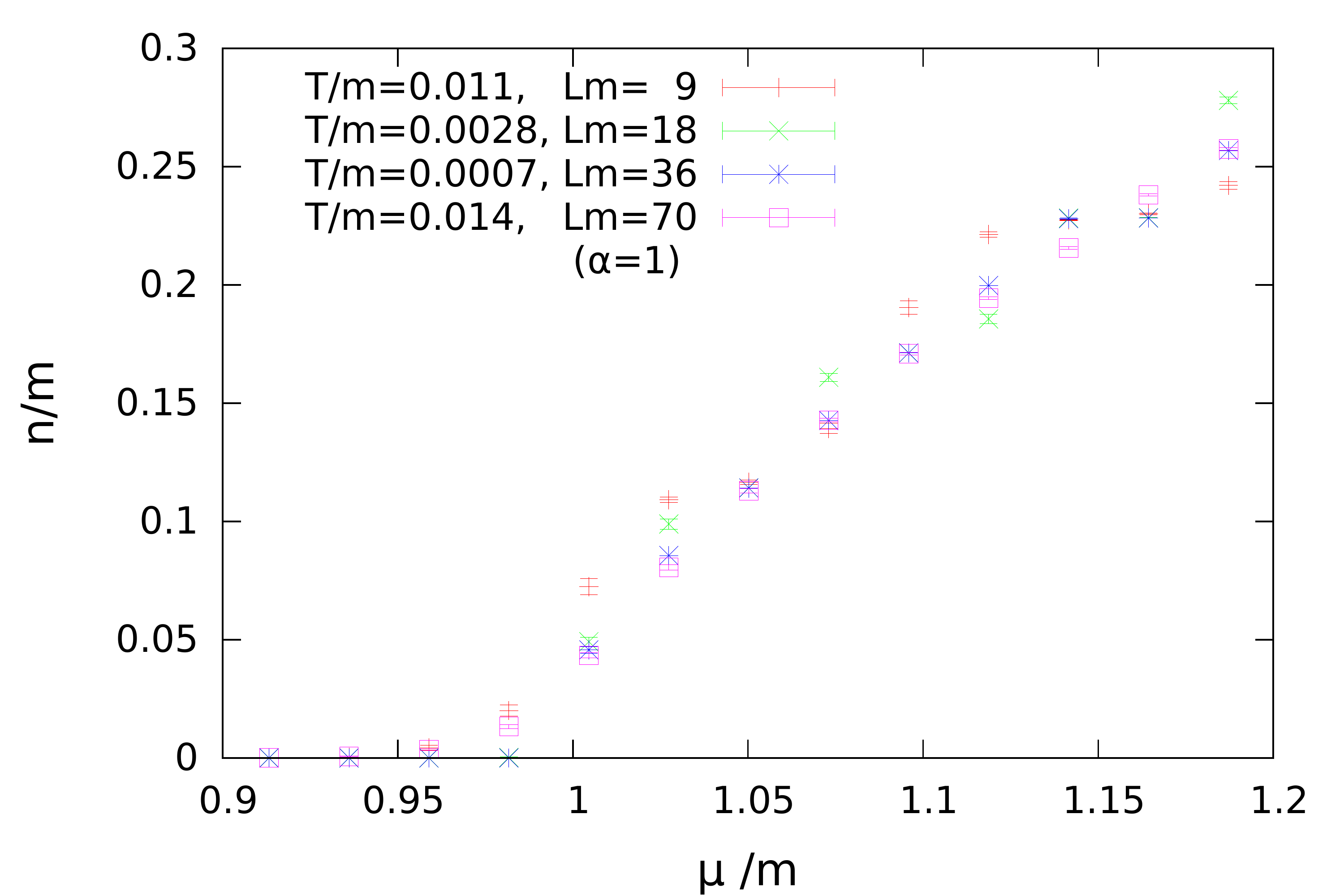}
 \caption{The density as a function of $\mu$ along the scaling trajectory $\alpha=1$ (left, where $T=1/L\to 0$) and $\alpha=2$ (right, where $T/m=0.88/(Lm)^2\to 0$, including the lowest $T$ data from the left panel for comparison).}
 \label{fig:vier}
\end{figure}

Fig.~\ref{fig:vier} shows the density for three lattices on these trajectories. Again, it vanishes for $\mu<m$ in the zero $T$ limit, but as a bulk quantity is not too sensitive to the particular scaling trajectory. 

\begin{figure}[!b]
 \includegraphics[width=0.49\linewidth,type=pdf,ext=.pdf,read=.pdf]{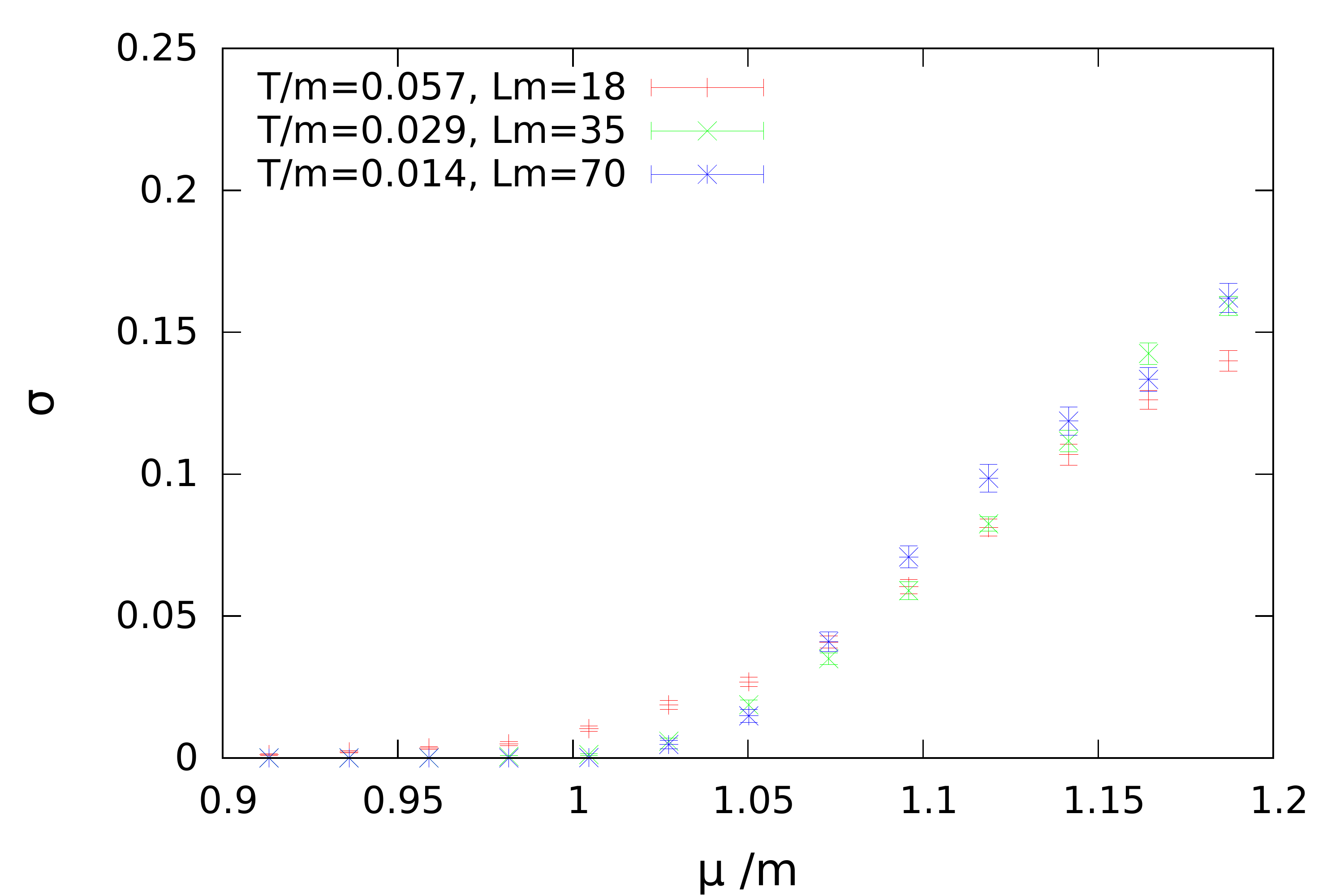}
 \hfill
 \includegraphics[width=0.49\linewidth,type=pdf,ext=.pdf,read=.pdf]{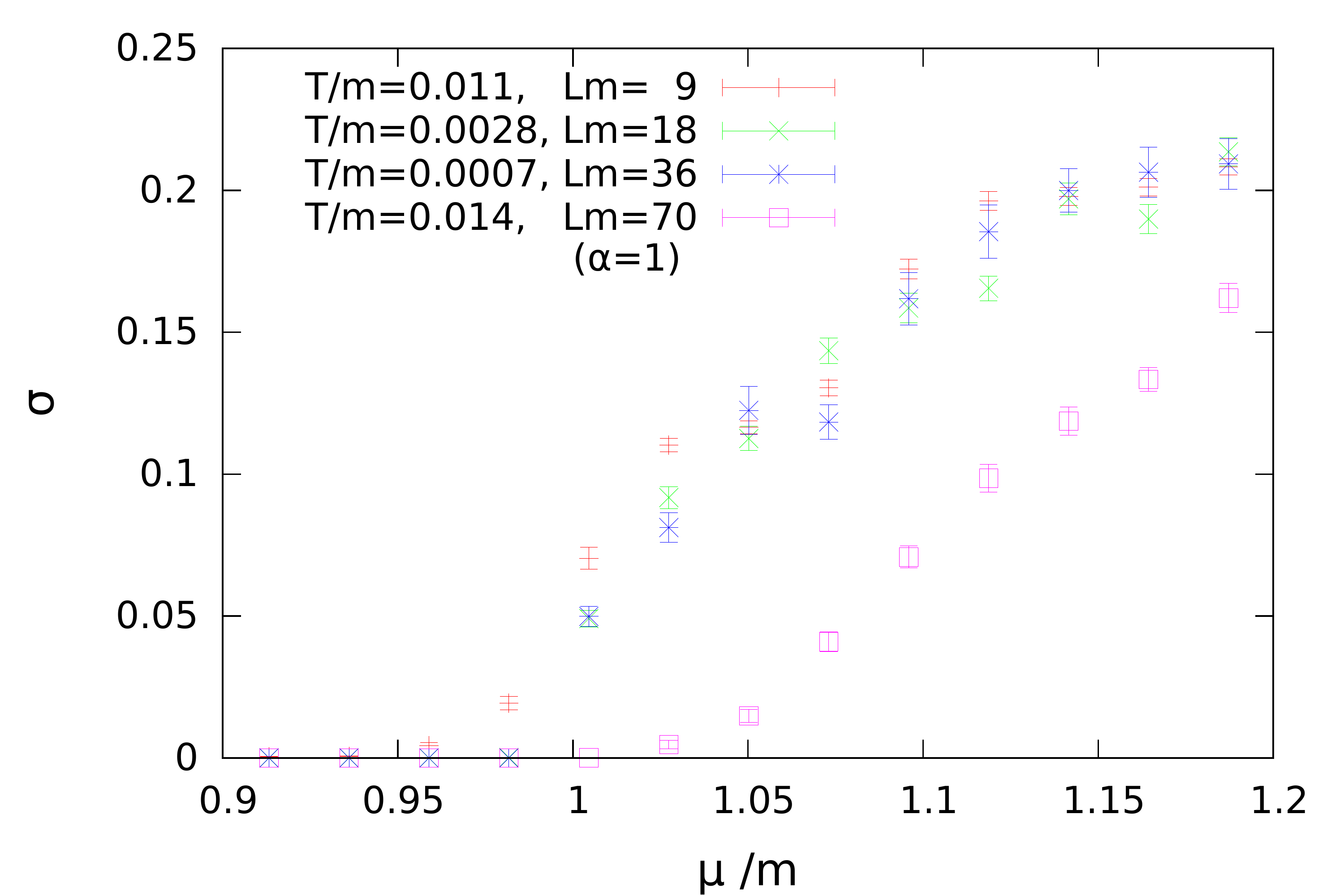}
 \caption{The spin stiffness, Eq.~\protect\eqref{eq:seven}, on the same scaling trajectories $\alpha=1$ and $\alpha=2$ as in Fig.~\protect\ref{fig:vier}.}
 \label{fig:fuenf}
\end{figure}

A quantity sensitive to spatial correlations\footnote{and in the O(2) model also to vortex condensation \cite{Mudry}} is the \textit{spin stiffness} $\sigma$, also called superfluid density. It is defined as the leading cost in free energy upon twisting the spatial boundary conditions,
\begin{equation}
 \phi(x_0,x_1+L)\stackrel{!}{=}\phi(x_0,x_1)+\varphi~,\qquad 
 \sigma:=L\,
 \frac{\partial^2 F}{\partial\varphi^2}
 \big|_{\varphi=0}
 =-LT\,
 \frac{1}{Z}\,\frac{\partial^2 Z}{\partial\varphi^2}
 \big|_{\varphi=0}~,
 \label{eq:seven}
\end{equation}
using that $F$ and $Z$ are even in the twist $\varphi$. The stiffness can be viewed as the susceptibility with respect to an `imaginary spatial chemical potential' $i\varphi/L$ and, therefore, can be measured from spatial winding numbers. Clearly, the stiffness is nonzero when the two ends of the spatial box are correlated, be it for exponential decays with spatial correlation length comparable or larger than $L$ or be it for infinite-range, e.g., algebraic decay.

From Fig.~\ref{fig:fuenf} one can see two properties of the stiffness: first, it is qualitatively similar to the density (and for $\alpha=\infty$, i.e., in the ground state at finite volume, one can actually show that $\sigma=n/m$ \cite{Bruckmann:2016txt}). Secondly, it does depend on the scaling trajectory, being significantly lower for $\alpha=1$ where the spatial size $L$ diverges faster and spatial correlations can get lost.

For understanding the stiffness results, we recall that in this limit the system has two infinite directions and associated correlation lengths diverging in a second order transition. The correlation length $\xi_\tau$ in the Euclidean time may differ from the spatial correlation length $\xi$, in particular since the particle worldlines at nonzero density distinguish the two directions. The two are related by the \textit{dynamical critical exponent} $z$\footnote{$z$ is also used for describing correlations in the fictitious time of Monte-Carlo simulations.} \cite{Hohenberg:1977ym}, i.e.,
\begin{equation}
 \xi_\tau\sim \xi^z~.
\end{equation}
Note the similarity to our scaling trajectories $\beta:=1/T\sim L^\alpha$. In fact, from the finite size scaling of the free energy density  
\cite{Fisher:1989zza}, 
\begin{equation}
 f\sim\frac{1}{L\beta}\,g\big(\frac{\xi}{L},\frac{\xi^z}{\beta};\varphi\big)~,
\end{equation}   
with $g$ a universal function, and a corresponding scaling of the stiffness, using in addition $\xi\sim(\mu/m-1)^\nu$, we can test \cite{vanOtterlo:1995} proposed values of the coefficients $z$ and $\nu$. Our numerical data prefer $z=2$ and $\nu=1/2$ \cite{Bruckmann:2016txt}, again in agreement with one-dimensional fermions.

\end{document}